# Enhanced spin pumping efficiency in antiferromagnetic IrMn thin films around the magnetic phase transition

L. Frangou,[1,2,3] S. Oyarzun,[4,5] S. Auffret,[1,2,3] L. Vila,[4,5] S. Gambarelli,[6,7] and V. Baltz[1,2,3,*]

[1] *Univ. Grenoble Alpes, SPINTEC, F-38000 Grenoble, France*
[2] *CNRS, SPINTEC, F-38000 Grenoble, France*
[3] *CEA, INAC-SPINTEC, F-38000 Grenoble, France*
[4] *Univ. Grenoble Alpes, NM, F-38000 Grenoble, France*
[5] *CEA, INAC-NM, F-38000 Grenoble, France*
[6] *Univ. Grenoble Alpes, SCIB, F-38000 Grenoble, France*
[7] *CEA, INAC-SCIB, F-38000 Grenoble, France*
[*] *To whom correspondence should be addressed: vincent.baltz@cea.fr*

**Abstract**

We report measurement of a spin pumping effect due to fluctuations of the magnetic order of IrMn thin films. A precessing NiFe ferromagnet injected spins into IrMn spin sinks, and enhanced damping was observed around the IrMn magnetic phase transition. Our data was compared to a recently developed theory and converted into interfacial spin mixing conductance enhancements. By spotting the spin pumping peak, the thickness dependence of the IrMn critical temperature could be determined and the characteristic length for the spin-spin interactions was deduced.

In spintronic materials, spin currents are key to unravelling spin dependent transport phenomena [1]. Researchers have sought to generate spin currents, and the spin pumping effect has attracted considerable attention due to its versatility [2,3]. In these studies, a variety of properties were investigated, such as spin penetration lengths [4] and the inverse spin Hall effect [5]. Spin pumping is applicable with all kinds of materials and magnetic orders: ferromagnets [6], paramagnets [7], antiferromagnets [8], and spin-glasses [9]. In addition, this technique disregards the electrical state whether the material is metallic [10], semiconducting [11], or insulating [12]. Spin pumping results from the non-equilibrium magnetization dynamics of a spin injector, which pumps a spin current ($I_S$) into an adjacent layer, called the spin sink. This spin sink absorbs the current to an extent which depends on its spin-dependent properties [13]. In practice, magnetization dynamics is most often driven by ferromagnetic resonance. The properties of the spin sink properties can be recorded either through the changes induced in ferromagnetic damping (inset of Fig. 1) or through direct electrical means, such as by measuring the inverse spin Hall voltage [5]. Finally, to eliminate direct exchange interactions and focus only on the effects due to the interaction between the spin current and the spin sink, the injector and the sink can be separated by an efficient spin conductor, such as copper.

The initial theoretical framework of spin pumping involves adiabatic charge pumping and a quality called spin mixing conductance [13]. Interfacial spin mixing conductances can be considered to act as filters for the spin current, and predictions based on these assumptions produce very good agreement with experimental data [4,7,14]. More recently, a linear-response formalism was developed to complete the existing theories and describe spin pumping near thermal equilibrium [15]. This formalism predicts a large enhancement of spin pumping near the magnetic phase transition due to spin sink fluctuations. These predictions, if validated experimentally, would help us to progress towards the development of more efficient spin sources, while also providing an alternative method to probe magnetic phase transitions. This type of alternative method is particularly needed to deal with the case of materials with no net magnetic moments, such as antiferromagnets.

The recent emergence of a field of research called antiferromagnetic spintronics has renewed interest for antiferromagnetic materials. The antiferromagnetic order is resistant to perturbation by magnetic fields, produces no stray fields, displays ultrafast dynamics and may generate large magneto-transport effects. Several effects have already been investigated in antiferromagnetic materials, such as tunnel anisotropic magnetoresistance [16], anisotropic magnetoresistance [17], spin Seebeck [18], inverse spin Hall [10] and inverse spin galvanic effects [19]. In addition, the impact of ultrathin films on spin-orbit torques, like IrMn below 1 nm, is the subject of intense research, although transition temperatures have not been established for these systems [21]. Extrapolating for the case of all-antiferromagnets devices [20], the order-disorder Néel temperature would set the thermal threshold for data retention. This temperature relates to the exchange stiffness between antiferromagnetic moments [22,23]. Sometimes, it is mistakenly confused with the blocking temperature which is specific to ferromagnetic/antiferromagnetic exchange bias interaction, but the Néel temperature is intrinsic to the antiferromagnet [22,23]. The blocking temperature is easily determined experimentally, for example by measuring the loss of the hysteresis loop shift as the external temperature rises, or by using specific field-cooling protocols [24,25]. In contrast, it is much more challenging to determine the Néel temperature of an isolated antiferromagnetic thin film. Despite the importance of such a basic parameter for antiferromagnetic spintronics, very few quantitative data have been published so far because of a lack of routinely available rapid measurement techniques compatible with most antiferromagnetic thin films. To our knowledge, neutron diffraction [26], magnetic susceptibility [27], nanocalorimetry [28], and resistivity measurements [29] are only appropriate for sufficiently thick single layers or for multiply repeated thinner layers.

In this work, we investigated the absorption of a spin current by IrMn thin films during their magnetic phase transition. Spin pumping experiments were performed at various temperatures on //NiFe8/Cu3/IrMn($t_{IrMn}$)/Al2 (nm) multilayers. The stacks were deposited at room temperature on thermally oxidised silicon substrates by dc-magnetron sputtering. A variable thicknesses of IrMn, $t_{IrMn}$, was deposited from an $Ir_{20}Mn_{80}$ target (at. %). The NiFe8 layer was deposited from a $Ni_{81}Fe_{19}$ target. To prevent oxidization in air, an Al2 cap was added which forms an AlOx protective film. This layer is known to have low spin current absorption properties. Because the spin diffusion length of copper is much longer than 3 nm, a Cu3 layer can eliminate exchange bias coupling without altering the spin propagation between the NiFe and IrMn layers. Series of ferromagnetic resonance spectra were recorded for temperatures (T) ranging between 10 and 300K, using a continuous wave electron paramagnetic resonance spectrometer operating at 9.6 GHz fitted with a dual-mode rectangular cavity. No significant heating or related spin Seebeck effect was expected across the sample [18,30,31]. A typical resonance spectrum is shown in Fig. 1(a). For each temperature the peak-to-peak linewidth ($\Delta H_{pp}$) was determined by fitting the spectrum to a Lorenzian derivative. The total Gilbert damping (α) was extracted from [4]:

$$\alpha(T) = \frac{\sqrt{3}|\gamma|}{2\omega}\left[\Delta H_{pp}(T) - \Delta H_0(300K)\right], \quad (1)$$

where γ is the gyromagnetic ratio and ω is the angular frequency. $\Delta H_0$ is the inhomogeneous broadening, which relates to spatial variations in the magnetic properties. This parameter can be determined from standard $\Delta H_{pp}$ vs. ω/2π plots by using a separate, room temperature, broadband coplanar waveguide with frequencies ranging between 10 and 22 GHz [4]. For our polycrystalline films, $\Delta H_0$ was small (e.g. for $t_{IrMn}$ = 0 and 1.5 nm, $\Delta H_{pp}$ = 3.3 and 4 mT and $\Delta H_0$ = 0.1 and 0.4 mT, respectively). For similar bare NiFe films [32] [32], $\Delta H_0$ was found to be a temperature-invariant parameter [32]. It is therefore reasonable to estimate that $\Delta H_0(T) = \Delta H_0(300K)$. Figure 1(b) shows α plotted against temperature. The data for $t_{IrMn}$ = 0 correspond to the temperature dependence of the local intrinsic NiFe Gilbert damping: $α^0(T)$. The signal shows superimposed decreasing and increasing components, qualitatively agreeing with the behaviour expected for typical 3*d* transition metals [33,34]. This can be readily understood based on predominant intraband scattering at low temperatures compared to the interband scattering prevailing at higher temperatures [34]. In the presence of the IrMn layer, the NiFe damping is the sum of local intrinsic damping and additional non-local damping ($α^p$) associated with the IrMn layer acting as a spin sink. From Fig. 1(b), we estimated $α(295K)$ ~ (8.5, 8.7, 9.9, 9.25, 9.75, and 9.75) x $10^{-3}$ for $t_{IrMn}$ = 0, 0.6, 0.8, 1, 1.2, and 1.5 nm, respectively. The overall increase of α with spin sink thickness up to a plateau from $t_{IrMn}$ = $λ_{IrMn}$ / 2, was extensively discussed in an earlier work describing room temperature measurements and larger thickness ranges [4]. This profile relates to the finite spin penetration length, $λ_{IrMn}$, for the spin sink. For IrMn, the penetration length at room temperature was approximately 0.7 nm [8]. The higher value we observed for $t_{IrMn}$ = 0.8 nm may be due to oscillations when nearing saturation [35,36], but this is beyond the scope of the present manuscript.

The temperature dependence of the IrMn contribution to NiFe damping can be directly isolated from: $α^p(T) = α(T) - α^0(T)$, as illustrated in Fig. 2(a). The central point of our article is that the signal displays a bump in $α^p$ ($δα^p$), highlighting a novel enhanced spin pumping effect. The position of this spin pumping peak depends on the temperature, which is related to the thickness of the IrMn layer. The effect is isolated in Fig. 2(b). In fact, a recent theory links $δα^p$ to the interfacial spin conductance [15]. This spin conductance depends on the dynamic transverse spin susceptibility of the spin sink, which is known to vary around critical temperatures. Transposed to our case:

$$\delta\alpha^p = \frac{1}{4\pi S_0 N_{SI}} g^{\uparrow\downarrow}_{Cu/IrMn}, \qquad (2)$$

where $S_0$ is the norm of the spin operator, $N_{SI}$ is the number of lattice sites in the NiFe spin injector (SI), and $g^{\uparrow\downarrow}_{Cu/IrMn}$ is the spin mixing conductance across the Cu/IrMn interface. As indicated by Ohnuma et al. [15], this last parameter is defined by:

$$g^{\uparrow\downarrow}_{Cu/IrMn} = \frac{8\pi J_{sd}^2 S_0^2 N_{int}}{\hbar^2 N_{SS}} \sum_k \frac{1}{\Omega_{rf}} \operatorname{Im}\chi_k^R(\Omega_{rf}), \qquad (3)$$

where $J_{sd}$ is the s-d exchange interaction at the Cu/IrMn interface, $N_{int}$ is the number of localized spins at the interface, $N_{SS}$ is the number of lattice sites in the IrMn spin sink (SS), $k$ is the wavevector, $\Omega_{rf}$ is the NiFe angular frequency at resonance, and $\chi_k^R(\Omega_{rf})$ is the dynamic transverse spin susceptibility of the IrMn layer. This model was initially developed for SS/SI bilayers but it can also be applied for the SS/Cu/SI trilayers described here since: i) spin absorption by 3 nm of Cu is negligible, and ii) the contribution of the SS/Cu interface is cancelled out when calculating $\alpha^p$.

Alternatively, the variation corresponding to $g^{\uparrow\downarrow}_{Cu/IrMn}/S$ can be calculated from:

$$\frac{\delta g^{\uparrow\downarrow}_{eff}}{S} = \frac{4\pi M_{S,NiFe} t_{NiFe}}{|\gamma|\hbar} \delta\alpha^p, \qquad (4)$$

where $g^{\uparrow\downarrow}_{eff}$ is the effective spin mixing conductance across the whole stack, $M_{S,NiFe}$ is the saturation magnetization of the NiFe layer, and $t_{NiFe}$ is its thickness. We measured the temperature dependence of $M_{S,NiFe}$ separately using a vibrating sample magnetometer. The results confirmed that, in the 10-300 K range, far from the NiFe Curie temperature, $M_{S,NiFe}$ remains virtually constant, at around 750 kA.m$^{-1}$. For the specific case of NiFe/Cu/SS trilayers, as shown by Ghosh et al. [4], because of cancellation of terms, $g^{\uparrow\downarrow}_{eff} \sim g^{\uparrow\downarrow}_{Cu/SS}$. More specifically, $1/g_{eff}^{\uparrow\downarrow} = 1/g_{NiFe/Cu}^{\uparrow\downarrow} - 1/g_{Sharvin,Cu}^{\uparrow\downarrow} + 1/g_{Cu/SS}^{\uparrow\downarrow}$, with $g_{NiFe/Cu}^{\uparrow\downarrow} \sim g_{Sharvin,Cu}^{\uparrow\downarrow} = 15$ nm$^{-2}$. We therefore took $\delta g_{eff}^{\uparrow\downarrow} = \delta g_{Cu/IrMn}^{\uparrow\downarrow}$ in Eq. (4). The resulting values are given on the right y-axis of Fig. 2(b). Note that the experimental framework may differ from the ideal theoretical one, since the IrMn structure and the Cu/IrMn interface are altered by species mixing and alloy formation [37]. In addition, the influence on $g_{Cu/IrMn}^{\uparrow\downarrow}$ of the non-trivial orientation of the IrMn moments with respect to the interface [38] almost certainly complicates the real picture. As illustrated in the inset of Fig. 2(b), δα$^p$ reduces under the effect of thermal activation, whereas the overall width relating to distributions of critical temperatures appears to increase. Of the parameters in Eqs. (2) and (3), only $\sum_k \frac{1}{\Omega_{rf}} \mathrm{Im}\,\chi_k^R(\Omega_{rf})$ significantly depends on temperature, increasing when the spin-flip relaxation time is shortened [15]. This would be expected to result in an increase of δα$^p$ with temperature, therefore the question of the thermal evolution of the peak shape remains open.

Figure 3(a) illustrates how the IrMn layer critical temperature ($T_{crit}^{IrMn}$) deduced from Fig. 2(b) is linearly related to its thickness. This behavior is corroborated by theoretical calculations taking magnetic phase transitions and finite size scaling into account [39]. The model considers the finite divergence of the spin-spin correlation length ($n_0$) near the critical temperature. For $t_{IrMn} < n_0$,

$$T_{crit}^{IrMn}(t_{IrMn}) = T_N^{IrMn}(bulk) \frac{t_{IrMn} - d}{2n_0}, \quad (5)$$

where $T_N^{IrMn}(bulk)$ is the Néel temperature of the IrMn bulk, equal to 700 K [26], and $d$ is the interatomic distance. X-ray diffraction measurements of similar samples revealed a (111) growth direction and a related interatomic distance, $d$, of about 0.22 nm, similar to that for bulk IrMn [26]. Fitting our data to Eq. 5 [Fig. 3(a)] returned a spin-spin correlation length of: $n_0 = 2.7$ +/- 0.1 nm (around 12 monolayers). Typical correlation lengths for ferromagnets range from a few monolayers up to ten monolayers [39]. The data point for $t_{IrMn}$ = 2 nm is taken from Petti et al. [20], but was measured by calorimetry on a different stacking. The level of agreement is nevertheless satisfactory. We also noted that $T_{crit}^{IrMn}$ = 300K for $t_{IrMn}$ ~ 2.7 nm. Extrinsic damping due to IrMn spin sinks ($\alpha^p$) [8] and the amplitude of the inverse spin Hall effect (ISHE) in IrMn layers [10] were found to be invariant around $t_{IrMn}$ ~ 2.7 nm at 300K. Thus, $\alpha^p$ and ISHE are only mildly sensitive to the static magnetic ordering, but more so to the nature of the elements constituting the alloy. Due to fluctuations in the magnetic order, a bump is still expected at the threshold thickness [40]. The ISHE was reported to be independent of the magnetic order for Cr [41], but dependent for PtMn [42]. Finally, for $t_{IrMn} > n_0$ the model presented by Zhang et al. [39] gives:

$$T_{crit}^{IrMn}(t_{IrMn}) = T_N^{IrMn}(bulk)\left[1 - \left(\frac{n_0 + d}{2t_{IrMn}}\right)^\lambda\right], \qquad (6)$$

with $\lambda$ = 1. Knowing $n_0$, and using Eq. (6), we can predict $T_N$ *vs* $t_{IrMn}$ for thick IrMn layers, as illustrated in the inset of Fig. 3(a).

Since critical temperatures are strongly linked to the extension of spin-spin interactions, we investigated the effect of the environment surrounding the IrMn layer. We fabricated //NiFe8/Cu3/IrMn0.8/Cap2 (nm) multilayers using various materials for the capping layer such as Pt and Pd, which are known to polarize easily. This could have enhanced $n_0$ and consequently $T_{crit}^{IrMn}$, but $T_{crit}^{IrMn}$ remains unaffected by its environment [Fig. 3(b)].

In conclusion, the main contribution of this paper is the experimental evidence that enhanced spin pumping efficiency can truly be achieved by using a fluctuating spin sink around the transition temperature for its magnetic order. This finding corroborates a recent theory linking enhanced spin pumping into a fluctuating spin sink to the interfacial spin mixing conductance. This spin mixing conductance depends on the transverse spin susceptibility of the spin sink, which is known to vary around critical temperatures. Spin pumping efficiency could be ultimately enhanced by including other magnetic orders and materials, preferably with large spin-orbit coefficients since larger enhancements are expected in such cases [15]. Finally, by showing that it is possible to detect magnetic phase transitions by spin pumping, this work also opens a new pathway for the further investigation of non-trivial magnetic orders, such as antiferromagnetism, with no net magnetic moment and potentially large magneto-transport effects. This type of magnetic order is expected to have a high potential in next-generation spintronic applications, a field known as antiferromagnetic spintronics. For example, by spotting the spin pumping peak, we experimentally determined how the IrMn critical temperature depended on the thickness of this layer. This information provided access to a fundamental parameter (the characteristic length for spin-spin interactions) which can be used to predict the full critical temperature *vs.* thickness dependence. Until now, for IrMn, this parameter had been experimentally inaccessible, and it remains to be measured for numerous common antiferromagnets, including FeMn, PtMn, and $Mn_2Au$.

After the initial submission of this manuscript we became aware of similar results for insulating NiO and CoO antiferromagnets obtained by Z. Qiu et al. [43]. These authors showed that, for YIG/NiO,CoO/Pt trilayers, the signature of the enhanced spin pumping efficiency at the magnetic phase transition could be detected through measurement of the inverse spin Hall voltage. Like $\alpha^p$, this voltage is linked to the interfacial spin mixing conductance.

**Acknowledgments**

We thank W. E. Bailey, A. Manchon, S. Maekawa, G. Gaudin, O. Klein, U. Ebels, and C. Hammel for valuable scientific discussions. We also thank J. Lopes for critical reading of the manuscript and M. Gallagher-Gambarelli of TWS editing for providing advice on English usage. We acknowledge the financial support of the French National Agency for Research (ANR) [Grant 'JCJC-ASTRONICS'].

**Figure captions**

Fig. 1. (color online) (a) Typical differential absorption spectrum at resonance. Inset: diagram representing the spin pumping experiment. (b) Dependence of α on temperature. The dashed lines are visual guides.

Fig. 2. (color online) (a) Dependence of $\alpha^p$ on temperature. To facilitate reading, the data were shifted vertically. Note: $\alpha^p$(295K) ~ (0.2, 1.4, 0.75, 1.25, and 1.25) x $10^{-3}$ for $t_{IrMn}$ = 0.6, 0.8, 1, 1.2, and 1.5 nm, which translates to $g_{eff}^{\uparrow\downarrow}/S$ (295K) ~ 0.8, 5.6, 3, 5, and 5 $nm^{-2}$, respectively. The baselines are visual guides. (b) Temperature dependence of $\delta\alpha^p$. To obtain $\delta\alpha^p$ vs. T, the baselines were removed from $\alpha^p$ vs. T. To facilitate reading, the data were multiplied by a factor of 3 for $t_{IrMn}$ = 1.5 nm. Inset: dependence of $\delta\alpha^p$ on temperature for T = $T_{crit}^{IrMn}$. An exponential function was fitted to the data as a visual guide.

Fig. 3. (color online) (a) Dependence of $T_{crit}^{IrMn}$ on $t_{IrMn}$. The line is a fit based on Zhang et al. [39], in the thin-layer regime. The data point for $t_{IrMn}$ = 2 nm is taken from Petti et al. [20]. Inset: $T_{crit}^{IrMn}$ vs $t_{IrMn}$ for a wider scale, along with the calculation in the thick-layer regime (dashed line). (b) $T_{crit}^{IrMn}$ for various capping layers.

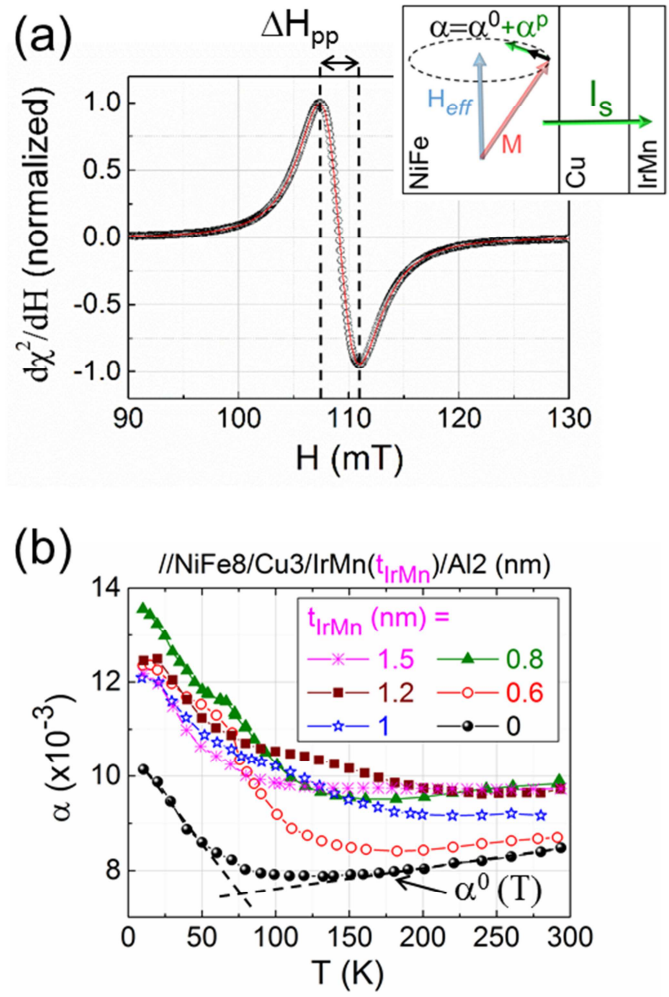

L. Frangou et al                                                                 Fig. 1

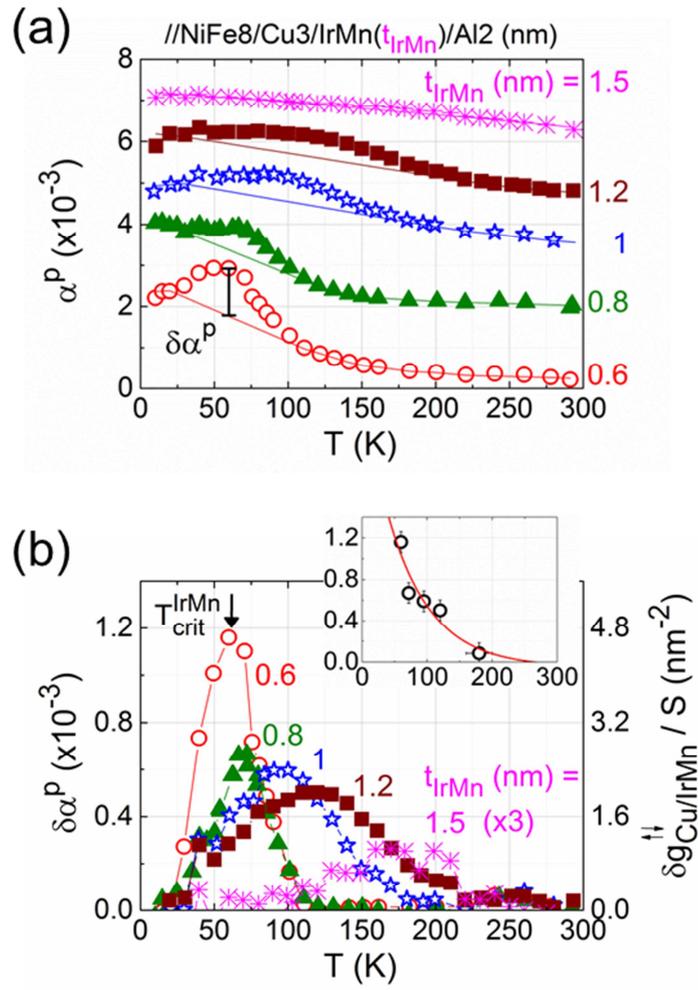

L. Frangou et al　　　　　　　　　　　　　　　　　　　　　　　　　　　　Fig. 2

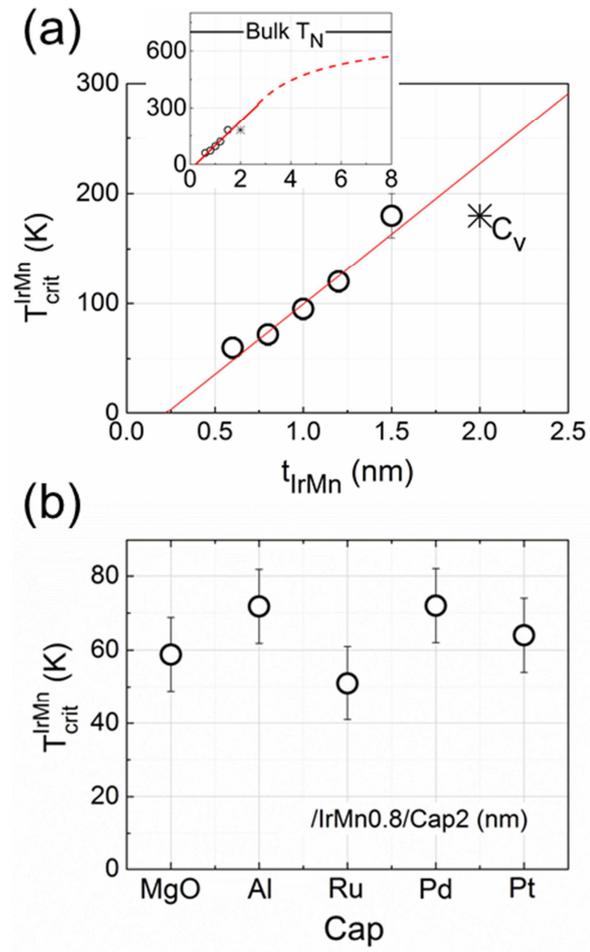